\documentclass[twocolumn,prl,showpacs,superscriptaddress,preprintnumbers]{revtex4}
\usepackage{amsmath,amssymb,epsfig,color}

\begin{document}

\title{Measurements of Raman scattering by electrons in metals: The effects
     of electron-phonon coupling}
\author{Yu. S. Ponosov}
\affiliation{Institute of Metal Physics UD RAS, 620990, S. Kovalevskaya str. 18, Ekaterinburg, Russia}
\author{ S. V. Streltsov}
\affiliation{Institute of Metal Physics UD RAS, 620990, S. Kovalevskaya str. 18, Ekaterinburg, Russia}
\affiliation{Ural Federal University, Mira str. 19, Ekaterinburg, Russia}

\begin{abstract}
We report the first systematic  measurements of the Raman scattering by electrons in elemental metals of 
Al, Mo, Nb, Os, Pb, Re, Ta, Ti, V, W and metallic compound La$B_6$. Experimental spectra are modelled on the base 
of the band structures, calculated within the density functional theory, taking properly into account the effects 
of electron-phonon scattering. The agreement between our measured and calculated spectra is excellent for the 
variety of metals, thus providing  the information on the electron self-energies and estimates for the electron-phonon coupling constants and temperature-dependent relaxation rates. The method can be applied for other metallic materials to evaluate an electron-phonon coupling as an alternative to the transport and optical measurements.
\end{abstract}   

\maketitle
 After the discovery of high-temperature superconductors (HTSC) the electronic Raman scattering (ERS) has been widely used to study electronic excitations both in superconducting and normal state of a variety 
 of compounds~\cite{1}. The ERS provides  information on the two-particle correlation function, which is in close correspondence to transport properties. In addition to studies of the magnitude and symmetry properties of the superconducting gap, many investigations of the ERS have been devoted to the problem of many-body interactions. Since clear understanding of the electron self-energy origin may shed light on the leading mechanism of unusual superconductivity, the ERS has become an important complementary tool for the electron-dynamics 
 investigations~\cite{1}. Thus, the observation of a flat, nearly frequency-independent, electronic Raman response in the normal state of 
 cuprates \cite{2,3,4,5,6} has become the basis of a marginal Fermi liquid 
 concept~\cite{7}. Models, incorporating different electron interactions, like nested Fermi liquid \cite{8} or antiferromagnetic spin 
 fluctuations \cite{9,10}, have been proposed to explain the unusual ERS spectra of HTSC. The phonon origin of the electron self-energy in cuprates is still 
 debated~\cite{11}. Broad Raman spectra very similar to a marginal Fermi liquid behavior have been simulated for strongly coupled 
 electron-phonon systems~\cite{12,13,14}. However, no experimental evidence for such behavior has been presented in early studies of the strongly-coupled conventional superconductors. Moreover, no normal-state spectra have been measured for 
 Nb,\cite{15,16} and the origin of the broad continuum in A-15 compounds, which is evolved in superconducting state, is still not fully 
 understood~\cite{15,16,17,18}. Thus, the investigations of the ERS spectra in the normal state of conventional metals and compounds, where an electron-phonon interaction seems to be the main contribution to the electron self-energy, and of their evolution upon the change of coupling strength and temperature are actually lacking.

The first unsuccessful attempts to observe the Raman scattering from electronic excitations in conventional metals were performed more than sixty years 
ago~\cite{19, 20}. The following calculation of the Raman response \cite{21} has shown that the sensitivity in these measurements was by 5 orders of magnitude low. In spite of the substantial advances of the Raman technique nowadays, the widespread opinion consist in that the measurements of the Raman effect from electrons in conventional metals is a difficult challenge. It is because the charge-density fluctuations are largely screened by conducting electrons and the q$\rightarrow$0 limit, usually used in the experiment treatment, implies the smallness of electronic scattering cross section which is proportional to $q^2$ at low frequencies. 

Most of elemental metals, excluding some alkali ones, have the anisotropic multisheeted Fermi surfaces. This leads to nonvanishing unscreened low-frequency  scattering \cite{22}. The increase of the effective q-vector  because of the strong absorption at a metallic surface \cite{23} (up to $2\times10^{6} cm^{-1}$  for the exciting laser energies $\omega_{i}$ in the visible range) also enhances the scattering intensity \cite{21,24}. With electron velocities $v_f$  being as high as $10^8$ cm/sec, the  electronic excitation's energies $\omega=qv_f$ in the  Raman spectra can spread up to 1000 $cm^{-1}$. Such large $qv_f$ values can not be neglected when treating the ERS in metals. The found resonance effects \cite{25} are another factors assisting the observation ERS spectra in metals.  

In this report we present the first systematic investigations of the temperature-dependent ERS spectra in the normal state of elemental metals and metallic compounds. The obtained spectra have been compared with those simulated based on the band structures, calculated within the density functional theory, taking into account the electron-phonon interaction. The groups of metals with weak and strong electron-phonon interaction were found to have strongly distinct Raman spectra and their temperature dependences. The results of the measurements and their analysis provided independent estimations of  the electron-phonon coupling constants and electron relaxation rates. 

 For temperature-dependent measurements (10-300K), electropolished plates of  single crystals (residual resistance ratio $\geq50$) were placed into an Oxford optical cryostat. The Raman spectra were excited by low-power laser radiation (up to 3 mW) at wavelengths of 514 nm and 633 nm, and they were recorded by a single-stage Renishaw microscope spectrometer providing a focal spot on the samples of 2 $\div$ 10 $\mu$m diameter. The represented spectra were corrected for the Bose factor, spectral response of the spectrometer, and transmission and absorption coefficients in the frequency range of the scattered light.

The cross section for the ERS by the intraband excitations is determined via integrating the electron susceptibility $\chi$(q, $\omega$) over the distribution of excited wave vectors 
$U^2(q)$ \cite{24, 26}:

\begin{equation}
\frac{d^2\sigma}{d\omega d\Omega}\propto\frac{1}{1-\exp^{-\hbar\omega/kT}}\int^{\infty}_{0}dq
\cdot U^{2}(q)\cdot \chi(q, \omega)
\end{equation}    
  When the interaction between electrons is taken into account, the susceptibility $\chi$(q, $\omega$) can be written in the form of the integral of the imaginary part of the q-dependent polarization operator over the Fermi surface [24, 26]:

\begin{eqnarray}
\chi_{\alpha,\beta}(q,\omega)=\oint \frac{ds_f}{\upsilon_f}
\left| \gamma_{\alpha,\beta}(k) \right|^2\int^{\infty}_{-\infty}
d\epsilon \left[ f(\epsilon)-f(\epsilon+\omega)
\right] \nonumber \\
\times\Im\frac{1}{\omega-q\upsilon_z-\Sigma^{'}
(\epsilon+\omega)+\Sigma^{'}(\epsilon)-i[\Sigma^{''}
(\epsilon+\omega)+\Sigma^{''}(\epsilon)]}
\end{eqnarray}   
The similar formalism has been previously used in the calculations of the nonadiabatic effects in the 
phonon self-energy\cite{27} and the ERS spectra in the  q$\rightarrow$0  limit~\cite{12,13,14}. 
Here, $f(\varepsilon)$ is the
Fermi function, and $z$ denotes the normal to the sample
surface. The electron velocities on the Fermi surface have been obtained within the band-structure calculations using the linearized muffin-tin orbital method (TB-LMTO-ASA)\cite{28}) in the local density approximation. Integration over the Fermi surface was performed 
with a fine mesh of 125,000 k-points in the full Brillouin zone.The retarded and advanced quasi-particle
electron self-energies $\Sigma(\epsilon)$ and
$\Sigma(\epsilon+\omega)$ determine the electron spectrum
renormalization near the Fermi level due to different
interactions. In the case of electron-phonon scattering,
their real and imaginary parts are ~\cite{29}

\begin{eqnarray}
\Sigma^{'}(\omega)= \int d\Omega\alpha^2F(\Omega){} \nonumber \\
{}\times \Re \left[ \psi \left( \frac{1}{2}+i\frac{\omega+\Omega}{2T}\right)-
\psi\left(\frac{1}{2}+
i\frac{\omega-\Omega}{2T} \right) \right]
\end{eqnarray}  
\begin{eqnarray}
\Sigma^{''}(\omega)= \pi \int d\Omega\alpha^2F(\Omega) \nonumber\\
\times \left[ 2n_B(\Omega)-f(\omega-\Omega)+f(\omega+\Omega) +1\right] +\nu 
\end{eqnarray}

The impurity relaxation rate $\nu$ was taken equal 10 $cm^{-1}$ in all calculations (this value is the upper limit for used pure samples), $\Psi$ is
the digamma function, $\Omega$ is the phonon energy,
$\alpha^2F(\Omega)$ is the Eliashberg spectral function for
the electron-phonon interaction, and $n_{\rm B}(\Omega)$ is the Bose
function. The literature data on the phonon densities of states $F(\Omega)$ \cite{30} and optical constants n and k 
\cite{31} for the calculation of the function $U^2$(q)  were used. Following \cite{32} we take $U^2$(q) to have the form
$\left|U(q)\right|^{2}\propto4\left|q\right|^{2}/\left|q^{2}-\xi^{2}\right|^{2}$, where 
$\xi=\xi_{1}-i\xi_{2}=(2\omega_{i}/c)\times(n-ik)$.  $U^2$(q) respresents a skew lineshape with a peak position $q_{0}=(2\omega_{i}/c)\times\sqrt{n^{2}+k^{2}}$. 

 Generally,  the matrix element of the electron-photon 
interaction $\gamma_{\alpha\beta}(k)$ in (2) ($\alpha$ and $\beta$ denote the polarizations for the
incident and scattered lights) includes both nonresonant (intraband) and resonant (interband) transitions~\cite{1,15,26}. Widely used for $\omega_{i}<<E_{g}$ effective-mass approximation  reduces $\gamma_{\alpha\beta}(k)$ to the Fermi surface curvature. In this case the account of the Coulomb interaction for scalar components induces  in (2) $\bar{\gamma}_{\alpha\beta}(k,q,\omega)=\gamma_{\alpha\beta}(k)-\left\langle 
\gamma_{\alpha\beta}(k,q,\omega)\right\rangle$ instead of bare $\gamma_{\alpha\beta}(k)$ \cite{22,24}; angular brackets denote the averaging over the Fermi surface. Indeed, the resonant terms in Raman vertex $\gamma_{\alpha\beta}(k)$ are important in metals, as it follows from the band structure calculations and was confirmed by our results ~\cite{25}. The calculation of the full Raman vertex requires a lot of computations, therefore,  to make them less cumbersome, we use Eq.(2) with constant $\gamma_{\alpha\beta}(k)$. This suggests the constancy of the momentum matrix elements and photon energy denominators in k space for the interband transitions.  Since the main aim is to analyze the ERS lineshapes, this rough simplification is partly justified by the similarity of the shapes for the spectra measured in different polarization geometries . This implies a slight effect of screening on the frequency dependence of the Raman response for the only screened $A_{1g}$ symmetry. It was shown in \cite{14,15} that the Coulomb interaction has to be taken into account only for the intraband processes. 
\begin{figure}[t]
\includegraphics[width=0.4\textwidth]{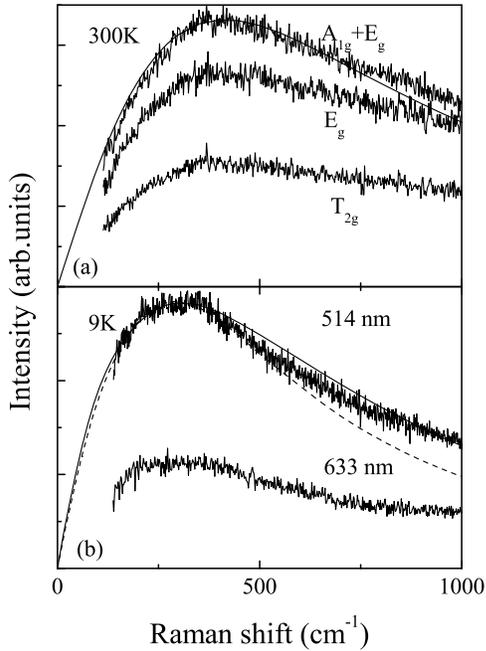}
\caption{\label{label}Raman spectra for W, measured in different polarization geometries with 514 nm excitation at 300K -(a) and using different exciting energies at 9K -(b). Calculated spectra are shown by solid line for $\lambda$=0.13 and dashed-for $\lambda$=0.}
\end{figure}

At first sight, all spectra, shown in Figs.1-3, look very similar. They contain broad continua with linearly 
increased intensity at small frequencies and a characteristic position of maximum. The spectra measured with different 
excitation energies have similar forms, as it is shown for W in Fig.1(b), where broad peaks with the energy of 
maximum  near 300 $cm^{-1}$ are observed at the low temperature for both excitating laser lines - 514 and 633 nm. 
The first-order Raman scattering by phonons for all investigated elemental cubic metals is forbidden and the upper 
limit of possible second-order spectra in W is limited by 400 $cm^{-1}$. Hence, the broad spectra observed indicate 
their electronic origin. The continuum frequency increases to 400 $cm^{-1}$ at room temperature. We calculated the 
ERS spectra for W using (Eq.2) with the electron-phonon constant $\lambda= 2\,\int d \Omega \alpha^2 F(\Omega)/\Omega=0$. In such case of noninteracting electrons the spectra maximum should be found at the energy $\omega=q \upsilon_z$, which is determined by the average Fermi velocity $\upsilon_z$ and wave vector q, corresponding to the maximum in the momentum transfer distribution  $U^2(q)$. One can see (dashed line in Fig.1(b)) that the energy and lineshape of the calculated low-temperature continuum are very close to the experimental ones. Such a coincidence evidences that the source of the observed ERS is the intraband electronic transitions near the Fermi level within conduction bands with bare dispersion. This also confirms the conservation of the electron momentum transfer q in the course of the scattering process that has already been  stated for Os \cite{25, 33}, where q-value variation leads to the continuum energy shift. Together with the moderate hardening of the continuum energy due to the increase of the quasiparticle damping upon increasing temperature, good agreement between experiment and calculation evidences rather weak electron interactions, i.e. small deviations from the Fermi liquid behavior. In fact, both low- and room-temperature spectra can be fitted well using $\lambda$=0.13 (Fig.1).  Obviously, in the case of W the $q \upsilon_z$ term can not be neglected even at room temperature. The shapes of the observed continua are quite similar for different polarization geometries ($A_{1g}$+$E_{g}$), $E_{g}$ and $T_{2g}$ (Fig.1(a)) indicating the same quasiparticle lifetimes for different symmetry channels.

\begin{figure}[t]
\includegraphics[width=0.4\textwidth]{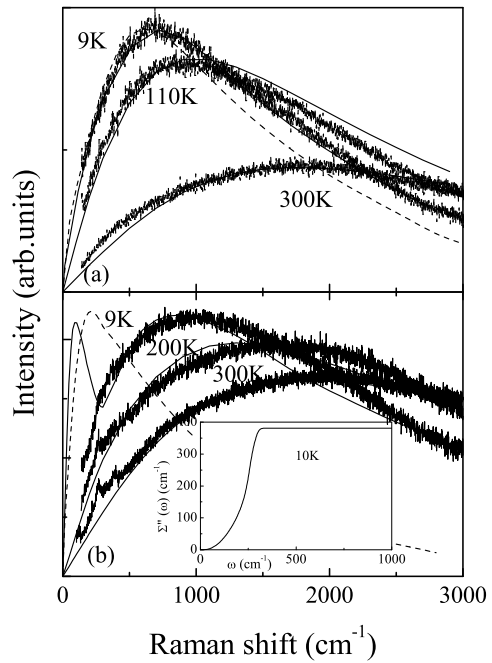}
\caption{\label{label}Raman spectra for Pb (a) and Nb (b) at different temperatures, measured with 633 nm excitation. Calculated spectra are shown by solid line for $\lambda$=1.04 and dashed-for $\lambda$=0. Inset in (b) - energy dependence of the quasiparticle damping for Nb at 10K, calculated with Eq. 4.}
\end{figure}

Next case when the $q\upsilon_z$ term determines the continua energies in the ERS spectra is shown in 
Fig.2(a). Though Pb is a strongly coupled superconductor, its low-temperature ERS spectrum is very 
close to the calculated spectrum for noninteracting electrons (dashed line in Fig.2(a). This is due to a large 
value of the $q\upsilon_z$ term (both Fermi velocity and q are large) and the smallness of the phonon-induced 
electron self-energies at low temperatures. The effect of temperature on the quasiparticle damping is large: the fit 
of the temperature-dependent Raman spectra with Eq.2 shows the increase of the electron relaxation rate 
$\Gamma\approx2\cdot \Sigma ''\left( \omega  \right)$ from 150 $cm^{-1}$ at l0K to 1500 $cm^{-1}$ at room 
temperature. At high temperatures this rate becomes frequency-independent and exceeds the contribution of 
the term with $q\upsilon_z\approx$600 $cm^{-1}$. In this case, the $q\upsilon_z$ term can be neglected and a 
Drude-like expression for the electron response $\chi(q, \omega)$ can be used in the $q \rightarrow 0$  limit [34]:

\begin{equation}
\chi(\omega)\propto N_f\frac{\omega\Gamma(\omega)}{\omega^2+\Gamma(\omega)^2}
\end{equation} 

where $N_f$ is the density of states at the Fermi level. The continuum energy then should be observed at the frequency $\omega\approx\Gamma$, which accounts for the large shift of the ERS spectra toward higher energies when temperature increases. The experimental spectra at all temperatures were well reproduced in our calculations with $\lambda$=1.04 (Fig.2(a)).   As in case of W, only small part of the spectral intensity in Pb is redistributed to the high-energy range at low temperatures, indicating  small deviations from the Fermi-liquid behavior.  At high temperatures, a large incoherent part appears over a high-energy scale that reminds the situation in HTSCs.

In another strongly coupled superconductor Nb non-Fermi liquid behavior can be observed already at low temperatures. While the position of maximum in the calculated ERS spectra for noninteracting electrons is 210 $cm^{-1}$,  the experimental value $\omega\approx$ 1000 $cm^{-1}$, i.e. five times larger (Fig.2(b)). For Nb, both the real and imaginary parts of the electron self-energy have large values already at low temperatures. This results in the appearance of a low-frequency feature in our calculation (Fig.2(b)).  This quasiparticle peak arises due to a large mass renormalization and can be still observed if the quasiparticle damping is small in this energy range, as our calculation shows (inset in Fig.2(b)). The same result was found for other strongly coupled metals Ta and V. However, we did not observe  signatures of these features up to the detection limit of our spectrometer ($\approx$100 $cm^{-1}$). Possibly, this indicates an increase of coupling at low frequencies as compared with the calculation, which can lead to suppression of the low-frequency peaks. The large quasiparticle damping (inset in Fig.2(b)), which becomes frequency-independent at high temperatures, determines the appearance of broad relaxation peaks at higher energies at all temperatures, which were well reproduced in the calculations (Fig.2(b)). Thus, we firstly report the observation of the normal-state intraband electronic continuum in Nb, the redistribution of which has been evidenced 
earlier\cite{18}. The electron-phonon coupling constant in Nb was estimated to be  $\lambda$=1.27.

\begin{figure}[b]
\includegraphics[width=0.4\textwidth]{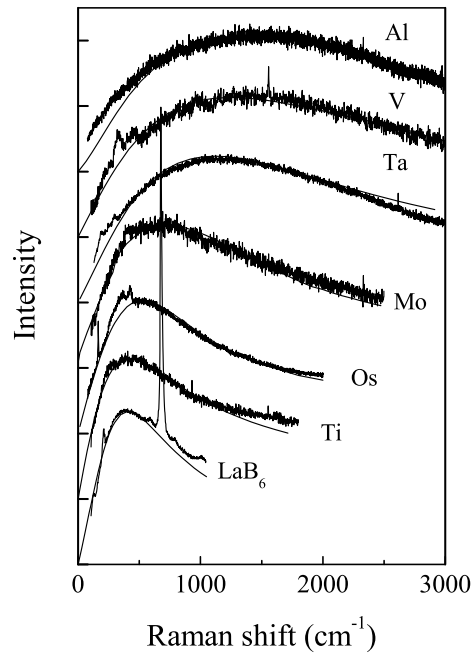}
\caption{\label{label}Raman spectra for a number of metals taken at 300K. Narrow lines in some spectra are the first- and second-order phonon scattering. Calculated spectra are shown by solid lines. The broad bump near 2000 $cm^{-1}$ in La$B_{6}$, observed only with 514 nm excitation, is not the Raman peak.}
\end{figure} 

The room temperature spectra for a number of metals are shown in Fig.3. As in the abovementioned examples the spectra contain the broad continua sometimes with superimposed peaks arising from the first- and second-order phonon scattering. All spectra show rather strong temperature dependences which will be reported elsewhere. The comparison with the calculation helps to explain the energy positions of the experimental peaks for each metal. For example, the ERS spectra for Al and V looks very similar but their continua energies are governed by different processes. In Al the spectrum position is determined by the $q\upsilon_z$ term  because its room-temperature relaxation rate $\Gamma\approx$400 $cm^{-1}$ is much lower than q$\upsilon_z\approx$1200 $cm^{-1}$. For V the situation is quite contrary and the spectral weight is transferred to high energies due the strong electron-phonon scattering. The frequency dependences of the ERS peaks and their temperature dependences are satisfactorily fitted by (Eq.2) and magnitudes of the electron-coupling constant $\lambda$ are estimated from these fits (Fig.3). The derived values of $\lambda$ and relaxation rates $\Gamma$ are given in Table I for all investigated metals together with available data from literature. 
We should note a rather good coincidence between the derived and literature $\lambda$ for strongly coupled metals (having superconducting temperatures $T_{c}\geq1^{\circ}K$) while for the low $T_{c}$ metals, our constants are, as a rule, smaller than the literature ones. The frequency dependence of the ERS spectra results from both the scattering processes of carriers due to interaction with phonons and Coulomb repulsion.  It may be suggested that the experiment measures the effective coupling constant which is renormalized by the Coulomb interaction.  It is such effective coupling constant that determines transition temperature for weakly coupled 
systems\cite{35}.
\begin{table}[t]
\caption{\label{tab:example}Experimental values for the electron-phonon coupling $\lambda_{exp}$ and room temperature relaxation frequency $\Gamma_{exp}$ (in $cm^{-1}$). Literature data of the transport 
$\lambda_{tr}$ \cite{36,37}, calculated by McMillan\cite{35,38} $\lambda_{McM}$  and $\Gamma_{opt}$ from optical 
experiments \cite{39,40,41} are shown for comparison.}
\begin{ruledtabular}
\begin{tabular}{llllll}
 & $\lambda_{exp}$&$\lambda_{tr}$&$\lambda_{McM}$&$\Gamma_{exp}$&$\Gamma_{opt}$\\
Al & 0.26&0.39&0.38&375&557\\
Mo & 0.33&0.32&0.41&450&446\\
Nb & 1.15&1.06&0.82&1500&1210\\
Os & 0.3&0.54&0.39&360&530\\
Pb & 1.04&1.48&1.12&1360&1625\\
Re & 0.77&0.76&0.46&1010&585\\
Ta & 0.83&0.87&0.65&1080&616\\
Ti & 0.31&0.54&0.38&430&400\\
V & 0.87&1.09&0.6&1140&637\\
W & 0.13&0.26&0.28&175&265\\
La$B_{6}$ &0.19&&0.33&270&160\\

\end{tabular}
\end{ruledtabular}
\end {table}

In summary, we presented convincing evidence for the ERS observation in metals, which is, indeed,  attributed to intraband electronic excitations. An account of finite q effects when calculating the ERS spectra is important, especially for metals with the weak electron-phonon coupling even at high temperatures. In the strongly coupled systems the relaxation rate for electrons usually overcomes the $q\upsilon_z$ term that allows one to obtain a fair description using the Drude-like expression in the collision limited regime. The electron-phonon interaction becomes apparent in the ERS spectra of metals owing to transfer of the spectral intensity to the high frequency range. It is found to be the leading channel of the electron self-energy in the investigated metals. The relaxation rates for electrons $\Gamma$ and electron-phonon coupling constants $\lambda$ are estimated from the comparison of the experimental spectra with the calculated ones. The agreement between the obtained  $\lambda$ values and those available in literature is fairly good. Thus, the ERS is shown to be an alternative to the transport and optical measurements to estimate the strength of the electron-phonon interaction.

 This research was partly supported by RFBR grants  No. 11-02-00306 and 10-02-96011 and by UD RAS program for young scientists. Authors acknowledge A.A. Eliseev for help with low temperature measurements.

\end{document}